\documentclass[submission,copyright,creativecommons]{eptcs}
\providecommand{\event}{FROM 2025}

\usepackage[T1]{fontenc}
\usepackage{cite}
\usepackage{amsmath,amssymb,amsfonts,bm}
\usepackage{url}
\usepackage{multicol}
\usepackage{textcomp}
\usepackage{xcolor}
\def\BibTeX{{\rm B\kern-.05em{\sc i\kern-.025em b}\kern-.08em
    T\kern-.1667em\lower.7ex\hbox{E}\kern-.125emX}}
    
\usepackage{graphicx}
\usepackage{listings}
\usepackage{mathtools}
\usepackage{tikz}
\usepackage{tikz-cd}
\usetikzlibrary{positioning}
\usetikzlibrary{calc}
\tikzset{stretch/.initial=1}
\usetikzlibrary{matrix,chains,positioning,decorations.pathreplacing,arrows}
\newcommand\drawloop[4][]%
   {\draw[shorten <=0pt, shorten >=0pt,#1]
      ($(#2)!\pgfkeysvalueof{/tikz/stretch}!(#2.#3)$)
      let \p1=($(#2.center)!\pgfkeysvalueof{/tikz/stretch}!(#2.north)-(#2)$),
          \n1= {veclen(\x1,\y1)*sin(0.5*(#4-#3))/sin(0.5*(180-#4+#3))}
      in arc [start angle={#3-90}, end angle={#4+90}, radius=\n1]%
   }

\usepackage{neuralnetwork}

\newcommand{\cgamma}{\gamma} 
\newcommand{\cf}[1]{\langle #1 \rangle} 

\usepackage{color}
\definecolor{keywordcolor}{rgb}{0.7, 0.1, 0.1}   
\definecolor{tacticcolor}{rgb}{0.0, 0.1, 0.6}    
\definecolor{commentcolor}{rgb}{0.4, 0.4, 0.4}   
\definecolor{symbolcolor}{rgb}{0.0, 0.1, 0.6}    
\definecolor{sortcolor}{rgb}{0.1, 0.5, 0.1}      
\definecolor{attributecolor}{rgb}{0.7, 0.1, 0.1} 

\lstdefinelanguage{Lean} {
    mathescape=false,
    texcl=false,
    morekeywords=[1]{
    import, prelude, protected, private, noncomputable, definition, meta, renaming,
    hiding, exposing, parameter, parameters, begin, conjecture, constant, constants,
    hypothesis, lemma, corollary, variable, variables, premise, premises, theory,
    print, theorem, proposition, example,
    open, as, export, override, axiom, axioms, inductive, with, without,
    structure, record, universe, universes,
    alias, help, precedence, reserve, declare_trace, add_key_equivalence,
    match, infix, infixl, infixr, notation, postfix, prefix, instance,
    eval, vm_eval, check, coercion, end, this, suppose,
    using, using_well_founded, namespace, section, fields,
    attribute, local, set_option, extends, include, omit, classes, class,
    instances, coercions, attributes, raw, replacing,
    calc, have, show, suffices, by, in, at, let, forall, Pi, fun,
    exists, if, dif, then, else, assume, obtain, from, aliases, register_simp_ext, unless, break, continue,
    mutual, do, def, run_cmd, const
    partial,mut,where,macro,syntax,deriving, return, try, catch, for, macro_rules,declare_syntax_cat,abbrev},
    morekeywords=[2]{Sort, Type, Prop},
    morekeywords=[3]{
    assumption,
    apply, intro, intros, allGoals,
    generalize, clear, revert, done, exact,
    refine, repeat, cases, rewrite, rw,
    simp, simp_all, contradiction,
    constructor, injection,
    induction,
    },
    literate=
    {->}{{\ensuremath{\mathrm{\to}}}}1
    {α}{{\ensuremath{\mathrm{\alpha}}}}1
    {β}{{\ensuremath{\mathrm{\beta}}}}1
    {γ}{{\ensuremath{\mathrm{\gamma}}}}1
    {Λ}{{\ensuremath{\mathrm{\Lambda}}}}1
    {δ}{{\ensuremath{\mathrm{\delta}}}}1
    {ε}{{\ensuremath{\mathrm{\varepsilon}}}}1
    {ζ}{{\ensuremath{\mathrm{\zeta}}}}1
    {η}{{\ensuremath{\mathrm{\eta}}}}1
    {θ}{{\ensuremath{\mathrm{\theta}}}}1
    {ι}{{\ensuremath{\mathrm{\iota}}}}1
    {κ}{{\ensuremath{\mathrm{\kappa}}}}1
    {μ}{{\ensuremath{\mathrm{\mu}}}}1
    {ν}{{\ensuremath{\mathrm{\nu}}}}1
    {ξ}{{\ensuremath{\mathrm{\xi}}}}1
    {π}{{\ensuremath{\mathrm{\mathnormal{\pi}}}}}1
    {ρ}{{\ensuremath{\mathrm{\rho}}}}1
    {σ}{{\ensuremath{\mathrm{\sigma}}}}1
    {τ}{{\ensuremath{\mathrm{\tau}}}}1
    {φ}{{\ensuremath{\mathrm{\varphi}}}}1
    {ϕ}{{\ensuremath{\mathrm{\varphi}}}}1
    {χ}{{\ensuremath{\mathrm{\chi}}}}1
    {ψ}{{\ensuremath{\mathrm{\psi}}}}1
    {ω}{{\ensuremath{\mathrm{\omega}}}}1
    {Γ}{{\ensuremath{\mathrm{\Gamma}}}}1
    {Δ}{{\ensuremath{\mathrm{\Delta}}}}1
    {Θ}{{\ensuremath{\mathrm{\Theta}}}}1
    {Λ}{{\ensuremath{\mathrm{\Lambda}}}}1
    {Σ}{{\ensuremath{\mathrm{\Sigma}}}}1
    {Φ}{{\ensuremath{\mathrm{\Phi}}}}1
    {Ξ}{{\ensuremath{\mathrm{\Xi}}}}1
    {Ψ}{{\ensuremath{\mathrm{\Psi}}}}1
    {Ω}{{\ensuremath{\mathrm{\Omega}}}}1
    {ℵ}{{\ensuremath{\aleph}}}1
    {≤}{{\ensuremath{\leq}}}1
    {≥}{{\ensuremath{\geq}}}1
    {≠}{{\ensuremath{\neq}}}1
    {≈}{{\ensuremath{\approx}}}1
    {≡}{{\ensuremath{\equiv}}}1
    {≃}{{\ensuremath{\simeq}}}1
    {≤}{{\ensuremath{\leq}}}1
    {≥}{{\ensuremath{\geq}}}1
    {∂}{{\ensuremath{\partial}}}1
    {∆}{{\ensuremath{\triangle}}}1 
    {∫}{{\ensuremath{\int}}}1
    {∑}{{\ensuremath{\mathrm{\Sigma}}}}1
    {Π}{{\ensuremath{\mathrm{\Pi}}}}1
    {⊥}{{\ensuremath{\perp}}}1
    {∞}{{\ensuremath{\infty}}}1
    {∂}{{\ensuremath{\partial}}}1
    {∓}{{\ensuremath{\mp}}}1
    {±}{{\ensuremath{\pm}}}1
    {×}{{\ensuremath{\times}}}1
    {⊕}{{\ensuremath{\oplus}}}1
    {⊙}{{\ensuremath{\odot}}}1
    {⊗}{{\ensuremath{\otimes}}}1
    {⊞}{{\ensuremath{\boxplus}}}1
    {□}{{\ensuremath{\Box}}}1
    {◇}{{\ensuremath{\Diamond}}}1
    {⊨}{{\ensuremath{\models}}}1
    {∇}{{\ensuremath{\nabla}}}1
    {√}{{\ensuremath{\sqrt}}}1
    {⬝}{{\ensuremath{\cdot}}}1
    {•}{{\ensuremath{\cdot}}}1
    {∘}{{\ensuremath{\circ}}}1
    {⁻}{{\ensuremath{^{-}}}}1
    {▸}{{\ensuremath{\blacktriangleright}}}1
    {∧}{{\ensuremath{\wedge}}}1
    {∨}{{\ensuremath{\vee}}}1
    {¬}{{\ensuremath{\neg}}}1
    {⊢}{{\ensuremath{\vdash}}}1
    {⟨}{{\ensuremath{\langle}}}1
    {⟩}{{\ensuremath{\rangle}}}1
    {↦}{{\ensuremath{\mapsto}}}1
    {←}{{\ensuremath{\leftarrow}}}1
    {<-}{{\ensuremath{\leftarrow}}}1
    {→}{{\ensuremath{\rightarrow}}}1
    {⊩}{{\ensuremath{\Vdash}}}1
    {↔}{{\ensuremath{\leftrightarrow}}}1
    {⇒}{{\ensuremath{\Rightarrow}}}1
    {⟹}{{\ensuremath{\Longrightarrow}}}1
    {⇐}{{\ensuremath{\Leftarrow}}}1
    {⟸}{{\ensuremath{\Longleftarrow}}}1
    {∩}{{\ensuremath{\cap}}}1
    {∪}{{\ensuremath{\cup}}}1
    {⊂}{{\ensuremath{\subseteq}}}1
    {⊆}{{\ensuremath{\subseteq}}}1
    {⊄}{{\ensuremath{\nsubseteq}}}1
    {⊈}{{\ensuremath{\nsubseteq}}}1
    {⊃}{{\ensuremath{\supset}}}1
    {⊇}{{\ensuremath{\supseteq}}}1
    {⊅}{{\ensuremath{\nsupseteq}}}1
    {⊉}{{\ensuremath{\nsupseteq}}}1
    {∈}{{\ensuremath{\in}}}1
    {∉}{{\ensuremath{\notin}}}1
    {∋}{{\ensuremath{\ni}}}1
    {∌}{{\ensuremath{\notni}}}1
    {∅}{{\ensuremath{\emptyset}}}1
    {∖}{{\ensuremath{\setminus}}}1
    {†}{{\ensuremath{\dag}}}1
    {ℕ}{{\ensuremath{\mathbb{N}}}}1
    {ℤ}{{\ensuremath{\mathbb{Z}}}}1
    {ℝ}{{\ensuremath{\mathbb{R}}}}1
    {ℚ}{{\ensuremath{\mathbb{Q}}}}1
    {ℂ}{{\ensuremath{\mathbb{C}}}}1
    {⌞}{{\ensuremath{\llcorner}}}1
    {⌟}{{\ensuremath{\lrcorner}}}1
    {⦃}{{\ensuremath{\{\!|}}}1
    {⦄}{{\ensuremath{|\!\}}}}1
    {‖}{{\ensuremath{\|}}}1
    {₁}{{\ensuremath{_1}}}1
    {₂}{{\ensuremath{_2}}}1
    {₃}{{\ensuremath{_3}}}1
    {₄}{{\ensuremath{_4}}}1
    {₅}{{\ensuremath{_5}}}1
    {₆}{{\ensuremath{_6}}}1
    {₇}{{\ensuremath{_7}}}1
    {₈}{{\ensuremath{_8}}}1
    {₉}{{\ensuremath{_9}}}1
    {₀}{{\ensuremath{_0}}}1
    {ᵢ}{{\ensuremath{_i}}}1
    {ⱼ}{{\ensuremath{_j}}}1
    {ₐ}{{\ensuremath{_a}}}1
    {¹}{{\ensuremath{^1}}}1
    {ₙ}{{\ensuremath{_n}}}1
    {ₘ}{{\ensuremath{_m}}}1
    {↑}{{\ensuremath{\uparrow}}}1
    {↓}{{\ensuremath{\downarrow}}}1
    {...}{{\ensuremath{\ldots}}}1
    {⟪}{{\ensuremath{\ll}}}1
    {⟫}{{\ensuremath{\gg}}}1
    {⋁}{{\ensuremath{\lor}}}1
    {⋀}{{\ensuremath{\land}}}1
    {·}{{\ensuremath{\cdot}}}1
    {▸}{{\ensuremath{\triangleright}}}1
    {Σ}{{\color{symbolcolor}\ensuremath{\Sigma}}}1
    {Π}{{\color{symbolcolor}\ensuremath{\Pi}}}1
    {∀}{{\color{symbolcolor}\ensuremath{\forall}}}1
    {∃}{{\color{symbolcolor}\ensuremath{\exists}}}1
    {λ}{{\color{symbolcolor}\ensuremath{\mathrm{\lambda}}}}1
    {\$}{{\color{symbolcolor}\$}}1
    {:=}{{\color{symbolcolor}:=}}1
    {=}{{\color{symbolcolor}=}}1
    {<|>}{{\color{symbolcolor}<|>}}1
    {<\$>}{{\color{symbolcolor}<\$>}}1
    {+}{{\color{symbolcolor}+}}1
    {*}{{\color{symbolcolor}*}}1,
    morecomment=[s][\color{commentcolor}]{/-}{-/},
    morecomment=[l][\itshape \color{commentcolor}]{--},
    showstringspaces=false,
    keepspaces=true,
    morestring=[b]",
    morestring=[d],
    tabsize=3,
    extendedchars=false,
    sensitive=true,
    breaklines=true,
    breakatwhitespace=true,
    basicstyle=\ttfamily\small,
    captionpos=b,
    columns=[l]fullflexible,
    identifierstyle={\ttfamily\color{black}},
    keywordstyle=[1]{\ttfamily\color{keywordcolor}},
    keywordstyle=[2]{\ttfamily\color{sortcolor}},
    keywordstyle=[3]{\ttfamily\color{tacticcolor}},
    keywordstyle=[4]{\ttfamily\color{attributecolor}},
    stringstyle=\ttfamily,
    commentstyle={\ttfamily\footnotesize },
}

\providecommand{\event}{SOS 2007}

\makeatletter
\newcommand\xLeftrightarrowceva[2][]{%
  \ext@arrow 9999{\longLeftrightarrowfill@}{#1}{#2}}
\newcommand\longLeftrightarrowfill@{%
  \arrowfill@\Leftarrow =\Rightarrow}
  \makeatother
\makeatletter

\DeclareRobustCommand\mos[1]{\mathrel{|}\joinrel
\stackrel{#1}{\mathrel{=}}}

\DeclareRobustCommand\vds[1]{\,\mathrel{|}\joinrel\joinrel\joinrel
\frac{#1}{ \ \ \ }}

\makeatletter

\newcommand\vso[2][]{%
  \ext@arrow 9999{\vsfill@}{#1}{#2}}
\newcommand\vsfill@{%
  \arrowfill@ \mid\ -}
\makeatother

\newcommand{\vp}{\varphi}

\newcommand{\mb}{\scriptscriptstyle{\Box}}

\newcommand{\N}{{\mathbb N}}
\newcommand{\R}{{\mathbb R}}
\newcommand{\Q}{{\mathbb Q}}

\newcommand{\nom}{N}

\newcommand{\ra}{\rightarrow}
\newcommand{\lra}{\leftrightarrow}
\newcommand{\su}{\texttt{\_}}

\newcommand{\bce}{\begin{center}}
\newcommand{\ece}{\end{center}}
\newcommand{\bcev}{\vspace*{-0.2cm}\begin{center}}
\newcommand{\ecev}{\end{center}\vspace*{-0.2cm}}

\newcommand{\bprop}{\begin{proposition}}
\newcommand{\eprop}{\end{proposition}}
\newcommand{\bcor}{\begin{corollary}}
\newcommand{\ecor}{\end{corollary}}

\newcommand{\RNum}[1]{\uppercase\expandafter{\romannumeral #1\relax}}

\newcommand{\ral}{\rightarrow_L}
\newcommand{\negl}{\neg_L}

\newcommand{\dia}{\Diamond}

\def\titlerunning{\L ukasiewicz Logic with Actions for Neural Networks training}
\def\authorrunning{ I. Leu\c stean \& B. Macovei}

\newcommand{\change}[1]{\textcolor{black}{#1}}
\usepackage{framed}
\providecommand{\urlalt}[2]{\href{#1}{#2}}
\providecommand{\doi}[1]{doi:\urlalt{http://dx.doi.org/#1}{#1}}
\begin{document}

\title{\L ukasiewicz Logic with Actions for Neural Networks training}

\author{Ioana Leu\c stean\textsuperscript{a} \and Bogdan Macovei\textsuperscript{a,b}}

\maketitle

\begin{center}
\textsuperscript{a}\,Faculty of Mathematics and Computer Science, University of Bucharest, Bucharest, Romania\\
\textsuperscript{b}\,Research Center for Logic, Optimization and Security (LOS), Department of Computer Science,
Faculty of Mathematics and Computer Science, University of Bucharest, Academiei 14, 010014 Bucharest, Romania\\[4pt]
\texttt{ioana.leustean@unibuc.ro}, \texttt{bogdan.macovei@unibuc.ro}
\end{center}

\def\titlerunning{\L ukasiewicz Logic with Actions for Neural Networks training}
\def\authorrunning{I. Leu\c stean \& B. Macovei}
\maketitle

\begin{abstract}
Based on  the already known connection between multilayer perceptrons and \L ukasiewicz logic with rational coefficients, we take a step forward in analyzing  its training process using a three-sorted hybrid modal logic:  a multilayer perceptron is a logical formula; the actions of the training process are modal operators; the training process is a sequence of logical deductions. Using the proof assistant and the programming language Lean 4, the algorithmic implementation of the training process is certified by logical proofs.

\end{abstract}

\section{Introduction}

The history of many-valued logics begins with the 3-valued system defined by J.  \L ukasiewicz in the twenties, which was extended to an $n$-valued system and to a $\infty$-valued system by J. \L ukasiewicz
and A. Tarski in 1930.

The $\infty$-valued \L ukasiewicz logic $Luk_\infty$ has the real interval $[0,1]$ as set of truth values. The logical connectives are  {\em implication} $\ra$ and {\em negation} $\neg$ with the   "truth-tables" defined by 
$\negl x= 1-x$ and $x \ral y = \min (1-x + y, 1)$ for any $x,y\in [0,1]$. The  axioms of the propositional calculus $Luk_\infty$ are:
\vspace*{-0.2cm}
\begin{center}
\begin{tabular}{llll}
(L1)& $\varphi\ral(\psi\ral\varphi)$ & (L3)& $(\varphi\ral\psi)\ral\psi)\ral(\psi\ral\varphi)\ral\varphi)$\\ 
(L2) &$(\varphi\ral\psi)\ral((\psi\ra\chi)\ral(\varphi\ral\chi))$ & (L4)& $(\neg\psi\ral\neg\varphi)\ral(\varphi\ral\psi)$
\end{tabular}
\end{center}
\vspace*{-0.2cm}
\noindent and the deduction rule is {\em modus ponens}. Note that adding  $((\varphi\ral\neg\varphi)\ral\neg\varphi)$ as an axiom, 
one gets the classical two-valued logic.

The connection between \L ukasiewicz logic and  neural networks is a consequence of the normal form theorem, proved in 1951 by R. McNaughton \cite{McN}: the term functions of \L ukasiewicz logic are $[0,1]$-valued, continuous, piecewise linear functions with integer coefficients defined on the unit cube. If the integer coefficients are replaced with  rational or real ones, then the corresponding theories are defined by adding a scalar multiplication to the traditional \L ukasiewicz operations.  These  normal forms are suitable for  representing  multilayer perceptrons with a particular activation function. Consequently, such a network can be defined by a logical formula in a natural extension of \L ukasiewicz logic. 

In this study we want to make a step forward and define the behaviour of a multilayer perceptron  as a logical inference.  To do this we use a three-valued hybrid modal logic with appropriate operators, based on the system ${\cal H}(@)$ defined in \cite{tableaux}. The  many-sorted hybrid modal logic as a general framework allows us to: (1) represent a multilayer perceptron as a logical formula; (2) represent the actions of the training process as modal operators; (3) represent the training process as logical deduction. To our knowledge, (2) and (3) represent a novel approach. Moreover, using "configurations" we increase the  versatility of our framework and we test it verifying a property of the training process. We use Lean 4 \cite{lean4}  as  proof assistant and as programming language,  the algorithmic implementation of the training process is certified by logical proofs.

In Section \ref{prel} we recall the results connecting \L ukasiewicz logic and neural networks. In Section \ref{luka} we recall  the general logic  ${\cal H}(@)$ and we define our particular system for representing and training multilayer perceptrons and we show that our system can be used  for verifying a property of the training process. 
The implementations in Lean 4 are briefly explained in  Section \ref{lean}\footnote {The complete implementation is available at:

 \url{https://github.com/bogdanmacovei/lukasiewicz-neural-network-formalization}}, which also contains a training of a real-world dataset.

\section{Preliminaries on \L ukasiewicz logic}\label{prel}
In this section we recall some basic facts about MV-algebras, their evolution and their connection with a class of neural networks. 

The algebraic structures corresponding to $Luk_\infty$ are the {\em MV-algebras},  defined by C.C. Chang in 1958 \cite{Chang1}:  an {\em MV-algebra} is a structure $(A,\oplus_L,\neg_L,0)$ of type (2,1,0)  such that $(A,\oplus_L,0)$ is an abelian monoid which satisfies the following properties for any $x,y \in A$:

\begin{tabular}{ll}
    (MV1) & $\neg_L (\neg_L x) = x$ \\
    (MV2) & $(\neg_L 0) \oplus_L x = \neg_L 0$ \\
    (MV3) & $\neg_L (\neg_L x \oplus_L y) \oplus_L y = \neg_L (\neg_L y \oplus_L x) \oplus_L x$
\end{tabular}

\noindent  MV-algebras stand to \L ukasiewicz $\infty$-valued logic as Boolean algebras stand to classical logic. 

\noindent In any MV-algebra $x\leq y$ if and only if $x\ral  y=1$. Moreover,  the {\em distance function} $d_L$ 
 is the usual distance on $[0,1]_L$, i.e. $d_L(x,y)=\mid x-y\mid$.
By Chang's completeness theorem \cite{Chang2}, the variety of MV-algebras is generated by the standard MV-algebra $[0,1]_L$. Consequently, the logic $Luk_\infty$ satisfies standard completeness (it is complete  with respect to evaluations in $[0,1]_L$).

In any MV-algebra $A$, we define the auxiliary operations for any $x,y \in A$:

\begin{tabular}{lll}
$1 := \neg_L 0$ &  $x \odot_L y {:=} \neg_L(\neg_Lx \oplus_L \neg_Ly)$ & $x\ral y := \neg_L x\oplus_L y$ \\
\change{$x \ominus y := x \odot \neg_L y$} & $x\vee_L y {:=} (x \odot_L \neg_L y) \oplus_L y$ &  $x \wedge_L y {:=} (x \oplus_L\neg_L y) \odot_L y$ \\
& $d_L(x,y):=(x\odot_L\neg_L y)\oplus_L (y\odot_L\neg_L x)$
\end{tabular}

\medskip 

Two notably examples of MV-algebras are:
\begin{itemize}
\item the Lindenbaum-Tarski algebra of $Luk_\infty$ is an MV-algebra with $[\vp]\oplus_L[\psi]=[\neg_L\vp\ral\psi]$, where 
$[\varphi]$  is the equivalence class of $\varphi$  with respect to 
$\vp\equiv \psi$ iff $\vdash\vp\ral\psi$ and $\vdash \psi\ral\vp$,
for any $\vp ,\psi$ formulas in $Luk_\infty$;
\item the standard algebra ${[0,1]_L}=([0,1],\oplus_L, \neg_L , 0)$ with \change{$x\oplus_L y := (x + y) \land 1$} for any $x,y\in [0,1]$.
\end{itemize}

 We refer to \cite{CDM} for all the unexplained notions related to MV-algebras and to \cite{MunBook} for advanced topics. The connection between \L ukasiewicz logic and neural networks is a consequence of two important results: the categorical equivalence between MV-algebras and abelian lattice-ordered groups and the normal form theorem  for
\L ukasiewicz logic. 

A lattice-ordered group \cite{BKW} is a group endowed with a lattice structure such that any group translation is isotone. 
D. Mundici proved in \cite{MundiciGr} that any MV-algebra is isomorphic with the unit interval of an abelian lattice-ordered group with strong unit. Consequently, the two classes of structures are categorically equivalent.  A natural problem was to extend this results for other classes of lattice-ordered structures (rings, vector spaces, algebras) and  MV-algebras endowed with additional operations.

A {\em Riesz MV-algebra} \cite{LeuRMV} is a structure 
$(R, \oplus, ^*,  \{r\mid r \in [0,1]\},0)$ such that $(R, \oplus, ^*, 0)$ is an MV-algebra and $\{r\mid r \in [0,1]\}$ is a family of unary operation such that the following  properties (RMV1)-(RMV4) hold:

\begin{tabular}{ll}
(RMV1) & $r (x\odot y^{*})=(r  x)\odot(r y)^{*}$ \\ 
(RMV2) & $(r\odot q^{*})\cdot x=(r x)\odot(qx)^{*}$ \\ 
(RMV3) & $r (q  x)=(rq) x$ \\
(RMV4) & $1x=x$.
\end{tabular}

 If we consider only rational scalars, i.e. $\{r\mid r \in [0,1]\cap\Q\}$, then we get an alternative characterization of  the class of  DMV-algebras (divisible MV-algebras). Riesz MV-algebras are categorically equivalent with vector lattices with strong unit, while DMV-algebras are categorically equivalent with vector lattices over $\Q$  \cite{DMVNotes}.  DMV-algebras and their logic were initially defined  and studied in \cite{Brunella}, using a family of operations $\{\delta_n\}_{n\in\N}$. The above axiomatization (RMV1)-(RMV4) is better suited for our purpose and it will be used in Section \ref{luka}.  
 
The varieties of Riesz MV-algebras and DMV-algebras are generated by the standard structure on $[0,1]$ and, respectively, $[0,1]\cap \Q$, the scalar multiplication being the real product. In this context, it is natural to seek characterizations for those functions $f:[0,1]^n\to [0,1]$  corresponding to logical formulas.  

The normal form theorem for \L ukasiewicz logic was firstly proved by R. McNaughton in 1951 \cite{McN}, a constructive proof being given in \cite{MuMN}: a function $f:[0,1]^n\to [0,1]$ is defined by a formula in \L ukasiewicz logic if and only if $f$ is a continuous piecewise linear function with integer coefficients. When \L ukasiewicz logic is endowed with rational (real) coefficients from $[0,1]$, the corresponding term function is a continuous piecewise linear function with rational (real) coefficients \cite{Brunella, LeuRMV}.

In \cite{Amato} the authors prove that a particular class of {\em multilayer perceptrons} coincide with the term functions of 
\L ukasiewicz logic with rational coefficients. A similar result for real coefficients is proved in \cite{wilf}.  More concretely,  a {\em multilayer perceptron}  with $k$ hidden layers, $n$ inputs and $n$ outputs can be represented as a function
$F:[0,1]^n \to [0,1]^n$  such that
$$y_j=
 \rho\left(\sum_{l=1}^{n} w^k_{jl}  
        \rho  \left(  \ldots\rho\left( \sum_{i=1}^{n} w^{0}_{pi}x_i + b^0\right)
    \ldots \right) + b_k\right),$$

\noindent where $F(x_1,\ldots,x_n)=(y_1,\ldots, y_n)$, $\rho:\R \to [0,1]$ is a
monotone-nondecreasing continuous function (referred to as
activation function) and  $w^k_{ij}$ are the synaptic weights in the $k^{th}$ layer and so on.  If $\rho(x)=ReLU_1(x):=min(1,max(0,x))$, then the functions corresponding to multilayer perceptrons coincide with the continuous piecewise  linear functions, so they can be represented as logical formulas in
\L ukasiewicz logic with appropriate coefficients. Moreover, in \cite{wilf} one can find an algorithm that calculates the formula corresponding to a linear combination from one layer.  

We  refer to \cite{Trillas} for an earlier connection between
\L ukasiewicz logic and neural networks and to \cite{lnn1, lnn2} for a recent one.

\section{A logic for neural networks}\label{luka}

Our  logic ${\mathcal H}_{\Sigma}(@)+\Lambda_{MLP}$ for specifying a multilayer perceptron and its training process is defined  in Section \ref{mlp}, as a particular theory of the many-sorted hybrid modal logic  ${\mathcal H}_{\Sigma}(@)$ \cite{tableaux}, which is briefly recalled in Section \ref{secHML}.

\subsection{The many-sorted Hybrid Modal Logic ${\mathcal H}_{ \Sigma}(@)$}\label{secHML}

For the many-sorted setting our reference is the system ${\mathcal H}_{\bm \Sigma}(@)$ as defined in \cite{tableaux}. Usually hybridization is performed on  top of a modal logic by adding nominals the satisfaction operator $@$ (more complex approaches are beyond the scope of the present paper).  Nominals allow us to directly refer the worlds (states) of a model, since they are evaluated in singletons in any model. However,  a nominal may refer to different worlds in different models. We refer to \cite{mod} for a  brief presentation of one sorted hybrid modal logic and to \cite{hybtemp} for a detailed one.

\begin{figure}[h]
\centering
\begin{framed}
{\small
\begin{itemize}
\item The axioms and the deduction rules of ${\mathcal K}_{\Sigma}$: 

\begin{itemize}
\item For any $s\in S$, if $\alpha$ is a formula of sort $s$ which is a theorem in propositional logic, then $\alpha$ is an axiom. 
\item Axiom schemes: for any $\sigma\in \Sigma_{s_1\cdots s_n,s}$ and for any formulas $\phi_1,\ldots, \phi_n,\phi,\chi$ of appropriate sorts, the following formulas are axioms:

\begin{tabular}{rl}
$(K_\sigma)$ & $\sigma^{\mb}(\ldots,\phi_{i-1},\phi\rightarrow\chi,\phi_{i+1}, \ldots)\to$\\ &\hspace*{0.5cm}$( \sigma^{\mb}(\ldots ,\phi_{i-1}, \phi, \phi_{i+1},\ldots) \to \sigma^{\mb}(\ldots ,\phi_{i-1}, \chi, \phi_{i+1},\ldots))$\\
$(Dual_\sigma)$& $\sigma (\psi_1,\ldots ,\psi_n )\leftrightarrow \neg \sigma^{\mb} (\neg \psi_1,\ldots ,\neg \psi_n )$
\end{tabular}

\item Deduction rules: {\em Modus Ponens} and {\em Universal Generalization}
\medskip

\begin{tabular}{rl}
$(MP)$ & if $\vds{s}\phi$ and $\vds{s}\phi\to \psi$ then 
$\vds{s}\psi$\\
$(UG)$ &  if $\vds{s_i}{\phi}$ then $\vds{s}\sigma^{\mb} (\phi_1, .. ,\phi, ..\phi_n)$
\end{tabular}
\end{itemize} 

\item Axiom schemes: any formula of the following form is an axiom, where $s,s',t$ are sorts,  $\sigma\in\Sigma_{s_1\cdots s_n,s}$,  $\phi,\psi, \phi_1,\ldots,\phi_n$ are formulas (when necessary, their sort is marked as a subscript), $j,k$ are nominals or constant nominals:

$\begin{array}{rlrl}
(K@) & @_j^{s} (\phi_t \to \psi_t) \to (@_j^s \phi \to @_j^s \psi) &
(Agree) &  @_k^{t}@_j^{t'} \phi_s \leftrightarrow @^t_j \phi_s\\
(SelfDual) & @^s_j \phi_t \leftrightarrow \neg @_j^s \neg \phi_t &
(Intro)  & j \to (\phi_s \leftrightarrow @_j^s \phi_s)\\
(Back) & \sigma(\ldots,\phi_{i-1}, @_j^{s_i} {\psi}_t,\phi_{i+1},\ldots)_s\to @_j^s {\psi}_t & 
(Ref) & @_j^sj_t \\
 (Nom\, x) & @_k x\wedge @_j x \to @_k j & & 
\end{array}$\\
\medskip

\item Deduction rules:

\begin{tabular}{rl}
$(BroadcastS)$ & if $\vds{s}@_j^s\phi_t$ then $\vds{s'}@_j^{s'}\phi_t$ \\
 $(Gen@)$& if $\vds{s'} \phi$ then $\vds{s} @_j \phi$, where $j $ and $\phi$ have the same sort $s'$\\
$(Name@)$ & if $\vds{s} @_l \phi$ then $\vds{s'} \phi$, where $l$ is a nominal\\
$(Paste)$ & if $\vds{s} @_j \sigma(\ldots, l,\ldots) \wedge @_l \phi \to \psi$ then $\vds{s} @_j \sigma(\ldots, \phi, \ldots) \to \psi$\\ & where $l$ is a nominal

\end{tabular}

Here, $j$ and $k$ are nominals or constant nominals having the appropriate sort. 
\end{itemize}
}
\end{framed}
\caption{{\bf The system} ${\mathcal H}_{\Sigma}(@)$
 \cite{tableaux}}
\label{fig:unu}
\end{figure}

A {\sf signature with constant nominals} is a triple $(S,\Sigma, {\nom})$ where $(S,\Sigma)$ is a many-sorted signature and ${\nom}=({\nom}_s)_{s\in S}$ is an $S$-sorted set of constant nominal symbols. In the sequel, we denote 
$\boldsymbol{ \Sigma}=(S,\Sigma, {\nom})$.

 For any $s\in S$ we define the formulas of sort $s$:
 \vspace*{-0.2cm}
\begin{center}
$\phi_s :=  p\mid j\mid y_s\mid \neg \phi_s \mid\phi_s \vee \phi_s \mid \sigma(\phi_{s_1}, \ldots, \phi_{s_n})_s
 \mid  @_k^s\phi_t$
\end{center}
\noindent Here, $p\in {\rm PROP}_s$, $j\in {\rm NOM}_s\cup {\nom}_s$, $t\in S$, $k\in {\rm NOM}_t\cup{\nom}_t$  and  $\sigma\in \Sigma_{s_1\cdots s_n,s}$. For any $\sigma\in \Sigma_{s_1\ldots s_,s}$, the dual formula is $\sigma^{\mb}(\phi_1,\ldots, \phi_n)=\neg \sigma(\neg \phi_1,\ldots, \neg \phi_n)$.

The deductive system is presented in Figure \ref{fig:unu}.

If \change{$\boldsymbol{ \Sigma}=(S,\Sigma, {\nom})$} then a $\boldsymbol{ \Sigma}$-{\sf frame} is ${\mathcal F}=((W_s)_{s\in S}, (R_\sigma)_{\sigma \in \Sigma}, ({\nom}_s^{\cal F})_{s\in S})$  such that    ${\nom}_{s}^{\cal F}=(w^c)_{c\in {\nom}_s}\subseteq W_s\neq \emptyset$ for any $s\in S$ and ${R}_{\sigma} \subseteq  W_s \times W_{s_1} \times \ldots \times W_{s_n}$  for any $\sigma \in \Sigma_{s_1 \cdots s_n,s}$.
We will further assume that distinct constant nominals have distinct sorts, so we shall simply write ${\nom}^{\cal F}=(w^c)_{c\in {\nom}}$. 
 Let \change{$\boldsymbol{\Sigma}=(S,\Sigma, {\nom})$} be a many-sorted signature with nominal constants and let  ${\cal F}$ be a $\boldsymbol{\Sigma}$-frame. A {\em model} (based on ${\cal F}$) is a pair
 ${\cal M}=({\cal F},V)$ such that $V:{\rm PROP}\cup {\rm NOM}\to {\cal P}(W)$ is an $S$-sorted function such that $V_s(j)$ is a singleton set for any $s\in S$ and $j\in  {\rm NOM}_s$.

For a model
 ${\mathcal M}=(W, (R_\sigma)_{\sigma\in\Sigma}, (w^c)_{c\in\nom}, V)$, $s\in S$, $w\in W_s$ and  $\phi$ a formula of sort $s$, the  many-sorted  \textit{satisfaction relation} $\mathcal{M},w\mos{s} \phi$
is inductively defined as usual.  We recall the definitions only for nominals and operators:
\begin{itemize}
\item $\mathcal{M},w \mos{s} j$, if and only if $ V_s(j)=\{w\}$ for any  $j\in {\rm NOM_s}\cup {\nom}_s$,

\item if $\sigma\in\Sigma_{s_1\ldots s_n,S}$ then $\mathcal{M},w \mos{s} \sigma(\phi_1, \ldots , \phi_n )$, if and only if there is \\$(w_1,\ldots,w_n) \in W_{s_1}\times\cdots\times W_{s_n}$ such that  $R_{\sigma} ww_1\ldots w_n$ and $\mathcal{M},w_i  \mos{s_i} \phi_i$ for any $i \in [n]$,
\item $\mathcal{M},w \mos{s} @_k^s\psi$ if and only if $\mathcal{M},u\mos{t} \psi$ where $k\in {\rm NOM}_t\cup {\nom}_t$,  $\psi$ has the sort $t$ and $V^{\nom}_t(k)=\{u\}$.
\end{itemize}

The completeness theorem is proved in \cite{tableaux}: the system ${\cal H}(@)$ is frame-complete and strong model-complete. Note that strong frame-completeness is proved for extensions  with pure formulas, i.e. formulas without propositional variables.

\subsection{${\mathcal H}_{\Sigma}(@)+\Lambda_{MLP}$}\label{mlp}
Even if the general theory, presented in Section \ref{prel} is developed both for  real and rational coefficients, in the sequel we  use  only  rational ones, so we denote
$[0,1]_\Q := [0,1]\cap \Q$.

In order to represent a multilayer perceptron and its training process is a three-sorted hybrid modal logic ${\mathcal H}_{\Sigma}(@)$ with  $\boldsymbol{\Sigma}=(S,\Sigma, N)$ and  $S=\{rmv,act,ln\}$. Since the general setting is presented in Section \ref{secHML}, we  only have to define the particular sets of operators and  constant nominals for each sort:  

\begin{tabular}{ll}  
$\Sigma_{rmv}=$ & $\{\negl: rmv\to rmv, \oplus_L: rmv\times rmv\to rmv\} \,\cup\, \{\dia_r :rmv\to rmv\mid r\in [0,1]_\Q\}$,\\[0.2cm]
$N_{rmv}=$& $\{\cgamma_r\mid r\in [0,1]_\Q\}$ is a set of nominal constants,\\[0.2cm]
 $\Sigma_{act}=$ &$ \{init, train, stop:rmv^{n}\to act \mid n\in \N\}$

 \\[0.2cm]
$\Sigma_{ln}=$& $\{[\su]\cf{\su} : act\times rmv^n \to ln \mid n\in \N\}$
\end{tabular}

\medskip

The formulas of our logic  are now completely defined. In the sequel we discuss the intended interpretation and the additional axioms for each sort.  

On $rmv$ we define the auxiliary operations $\odot_L$,  $\vee_l$, $\wedge_L$, $d_L$ as in  Section \ref{prel}. The additional axioms of sort  $rmv$ are presented in Figure \ref{fig:doi}.

\begin{figure}
\centering
\begin{framed}
{\small
\begin{itemize}
\item  Axioms  for nominal constants:

\begin{tabular}{llllll}
(Nom1) & $\cgamma_{\neg_L r}\lra \neg_L \cgamma_r$ & 
(Nom2) & $\cgamma_{r\oplus_L q}\lra \cgamma_{r}\oplus_L\cgamma_q$ & 
(Nom3) & $\cgamma_{r\cdot q}\lra \dia_r\cgamma_q$ 
\end{tabular}

\item  Axioms for the MV-algebraic operations:

\begin{tabular}{llll}
(M1) &$(\varphi\oplus_L(\psi\oplus_L \chi))\lra  (\varphi\oplus_L \psi)\oplus_L \chi$ &(M4) &$(\neg_l(\neg_L\varphi)) \lra (\varphi)$ \\ 
 (M2) &$((\neg_L \cgamma_0)\oplus_L\varphi) \lra (\neg_L\cgamma_0)$ & (M5)  &$(\varphi\oplus_L\psi) \lra (\psi\oplus_L\varphi)$\\
 (M3) & $((\varphi\odot_L\neg_L\psi)\oplus_L\psi) \lra ((\psi\odot_L\neg_L\varphi)\oplus_L\varphi)$ & (M6) &${\cgamma_0} \lra (\varphi\oplus_L\cgamma_0)$ 
\end{tabular}

\item Axioms for the scalar multiplication:

\begin{tabular}{llll}
(R1) &$(\dia_{r}(\varphi \odot_L\neg_L \psi))\lra   ((\dia_{r}\varphi) \odot_L \neg_L(\dia_{r}\psi))$ & (R4) &$(\dia_1\varphi)\lra \varphi$\\
(R2) &$(\dia_{r\odot \neg q} \varphi) \lra   
((\dia_{r}\varphi) \odot_L \neg_L(\dia_{q}\varphi))$ & 
(R3) &$(\dia_r(\dia q \varphi))\lra (\dia_{r\cdot q}\varphi)$ 
\end{tabular}
\end{itemize}
 where $r,q\in [0,1]_\Q$, $r\cdot q$ is the real product on $[0,1]$, $\lra$ is the modal equivalence from ${\mathcal H}_{\Sigma}(@)$ and $\varphi$, $\psi$, $\chi$ are arbitrary $rmv$-formulas.
}
\end{framed}
\caption{Axioms for $rmv$-formulas}
\label{fig:doi}
\end{figure}

\noindent If ${\mathcal M}=({\mathcal F},V)$ is a model,   then the set $W_{N}=\{V(\cgamma_r)\mid r\in [0,1]_{\Q}\}$, becomes a DMV-algebra if we define
 $\neg^W V(\cgamma) := V(\neg_L\cgamma)$, $V(\cgamma_1)\oplus^W V(\cgamma_2):= V(\cgamma_1\oplus_L\cgamma_2)$, $0^W:=V(0_L)$, and $r V(\cgamma_q):=V(\cgamma_{r\cdot q})$
 for any 
 $r,q\in [0,1]_\Q$. All the DMV-algebra operations are  well-defined by (Nom1), (Nom2) and (Nom3). 
 
  Note that, in our setting, the connectives of \L ukasiewicz logic with  rational coefficients are normal modal operators. Consequently, we can represent DMV-formulas and we can reason about them in a classical modal manner, within the general system  ${\mathcal H}_{\Sigma}(@)$. Moreover, using the constant nominals from $N_{rmv}$ we represent the rationals from $[0,1]$ as formulas in our logic.

  
  Given the above considerations, from now on, we shall denote
   $\cgamma_r$ by $\bf r$ for any $r\in [0,1]$, i.e.  ${\bf r}$ is a nominal constant of sort $rmv$. 
   
   In our presentation  we used the same letter for the worlds of a model $w\in W$ and the weights of the training process ${\bf w}$. Both are standard notation, one is a semantic notation, the other is a syntactic one, so we think there is no danger of confusion. Moreover, since the weights are actually represented as nominal constants, they correspond to a unique world in a natural way.

 In the sequel we assume that $n,k\in \N$. In order to define the training process of  a  multilayer perceptron  with $k$ hidden layers and  $n$ inputs  we make the following notations:
if $h=(h_1,\ldots, h_n)\in [0,1]_\Q^n$, we denote by ${\bf h}_1^n$ the vector $({\bf h}_1,\ldots, {\bf h}_n)$ of corresponding $rmv$-nominal constants; if $w=(w_{ij})_{i,j=1}^n\in M_n([0,1]_\Q)$
 is a square matrix then   we denote  by ${\bf w}:= ({\bf w}_{ij})_{i,j=1}^n$ the corresponding matrix of $rmv$-nominal constants; in general, for any set of elements $E$, we denote by $e_{i}^{i+j}$ a sequence $(e_i,\ldots, e_{i+j})\in E^j$. Moreover, when the notation is clear we remove the unnecessary brackets.

The atomic $act$-formulas are meant to define the actions: 
\begin{itemize}
    \item $train({\bf h}_1^n)$ performs a forward step for the $n$ inputs ${\bf h}_1^n$, 
 \item $init({\bf h}_1^n)$ starts the  forward training  for the $n$ inputs ${\bf h}_1^n$,
 \item $stop({\bf h}_1^n)$ stops the training process with the outputs ${\bf h}_1^n$; when the output is not relevant we simply use $stop()$ (see (N0$_E$) below). 
  \end{itemize}

The training process of a neural network is defined as an inference on the sort $ln$. The particular operators on this sort is $[\alpha_{act}]\cf{st_{rmv}}$ whose intended meaning is the following: $\alpha_{act}$ is an action and $st_{rmv}$  is a sequence of formulas of sort $rmv$ representing a configuration (or a state); the entire formula $[\alpha_{act}]\cf{st_{rmv}}$ means that in the state $st_{rmv}$ we perform the action $\alpha_{act}$. Note that we use $[\alpha_{act}]\cf{}$, which means that we reached the empty state and, in our setting, the continuation of the trainig process will be determined by analyzing the current input (see (N2) and (N3) below).   This is a departure from the well-known semantics of Propositional Dynamic Logic (PDL) \cite{pdl}, where $[\alpha]\varphi$ means that $\varphi$ holds {\em after} performing the action $\alpha$. Note that in our setting $\alpha_{act}$ is not interpreted as a transition between states: the semantic is the general one defined in Section \ref{secHML}.

We are ready to present the training axioms. 
Recall from Section \ref{prel} that our network is a multilayer perceptron, with the activation function $ReLU_1(x)=min(max(0,x),1)$, so the resulting values are determined using the algoritm from \cite{wilf}.

In order to define the axioms,  we make the following notations:

\begin{itemize}

\item[(n1)] $next_{{\bf w},{\bf b}}({\bf \lambda}_1^n):= ({\bf b}\oplus \bigoplus_{i=1}^n \dia_{w_{1i}}{\bf\lambda}_i,\ldots, {\bf b}\oplus \bigoplus_{i=1}^n \dia_{w_{ni}}{\bf\lambda}_i)$
\item[(n2)] $end({\bf y},{\bf \lambda}_1^n,\epsilon):=d_L({\bf y},  {\bigvee}_1^n{\bf \lambda}_i )\ral \epsilon$
\item[(n3)] $updated_{{\bf \lambda}_1^n}\,\, \cf{{\bf w}_0^k, {\bf b}_0^k}:= \cf{{\bf uw}_0^k, {\bf ub}_0^k} $.
\end{itemize}

\noindent where  ${\bf\lambda}_1^n$, ${\bf b}_0^k$  are vectors and ${\bf w}_0^k$ is a matrix of nominal terms of sort $rmv$. Note that $next$ defined the aggregation between consecutive layers  and $end$ defines the condition for ending the training, expressed as formulas in \L ukasiewicz logic. Note that $next$ uses the operations $\oplus$ in order to define a linear combination in \L ukasiewicz logic, while $end$ used the supremum $\bigvee$ in order to aggregate a value from $n$ outputs (but it can be replaced with other aggregation function). 

Given a configuration $\cf{{\bf w}_0^k, {\bf b}_0^k}$ and an output ${\bf \lambda}_1^n$, the new configuration obtained by performing backpropagation is determined using $updated$. An example is provided for the simple network from Figure \ref{fig:net}. We note that

 For a  neural  network with one  input
$(h_1,\ldots, h_n)\in [0,1]_\Q^n$  and the expected output $y\in [0,1]_\Q$ the axioms are:

\begin{itemize}

\item[(N0)]  $[init ({\bf h}_1^n)]\cf{{\bf w}_0^k, {\bf b}_0^k}\to [train ({\bf h}_1^n)]\cf{{\bf w}_0^k, {\bf b}_0^k}$\\

\item[(N1)] $[train ({\bf h}_1^n)]\cf{{\bf w}_i^k, {\bf b}_i^k}\to [train (next_{{\bf w}_i, {\bf b}_i}({\bf h}_1^n))]\cf{{\bf w}_{i+1}^k, {\bf b}_{i+1}^k }$\\

\item[(N2)] $[init ({\bf h}_1^n)]\cf{{\bf w}_0^k, {\bf b}_0^k} \to ([train(\lambda_1^n)]\cf{}\wedge \neg @^{ln}_{1_L}end({\bf y},  \lambda_1^n, {\bf \varepsilon})\to [init ({\bf h}_1^n)]
updated_{\lambda_1^n}\,\,  \cf{{\bf w}_0^k, {\bf b}_0^k})$\\

\item[(N3)] $[init ({\bf h}_1^n)]\cf{{\bf w}_0^k, {\bf b}_0^k} \to ([train(\lambda_1^n)]\cf{}\wedge  @^{ln}_{1_L}end({\bf y},  \lambda_1^n, {\bf \varepsilon})\to [stop (\lambda_1^n)]
\cf{{\bf w}_0^k, {\bf b}_0^k})$

\end{itemize}

\medskip

In \cite{tableaux} the authors analyze general logics ${\mathcal H}_{\bm \Sigma}(@)+\Lambda$, where $\Lambda$ is a particular set of axioms.   Our logic is ${\mathcal H}_{\bm \Sigma}(@)+\Lambda_{MLP}$, where 

$\Lambda_{MLP}= \{(Nom1)-(Nom3),(M1)-(M6), (R1)-(R4),(N(0)-(N3)\}$

\noindent Consequently, the  (weak) completeness results hold: our logic is complete with respect to the class of models defined by $\Lambda_{MLP}$.
\medskip

\medskip

As an example, we consider a simple neural network:

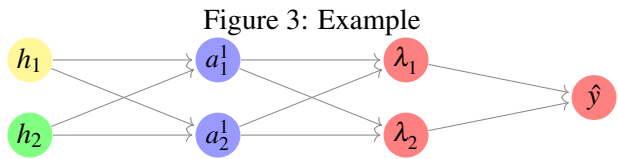
\begin{figure}[h]
\centering
\caption{Example}
\label{fig:net}
\begin{neuralnetwork}[height=2]
    \newcommand{\x}[2]{$h_{\number\numexpr#2+1}$}
    \newcommand{\y}[2]{$\hat{y}$}
    \newcommand{\hfirst}[2]{\small $a^{1}_#2$}
    \newcommand{\hsecond}[2]{\small $a^{2}_#2$}
    \newcommand{\lfirst}[2]{\small $\lambda^{}_#2$}
    \newcommand{\lsecond}[2]{\small $\lambda^{}_#2$}
    \inputlayer[count=1, bias=true, text=\x]
     \hiddenlayer[count=2, bias=false, text=\hfirst] \linklayers
   \outputlayer[count=2, bias=false, text=\lfirst] \linklayers    
   \outputlayer[count=1, text=\y] \linklayers
\end{neuralnetwork}
\end{figure}

\noindent where $n=2$, $k=1$, the inputs are $\bf{h} = (\bf {0.2},\bf{0.3})$, and the initial  weights and biases are: 
\begin{center}
 $\bf{w}_0= \left(\begin{array}{cc}
         {\bf 0.4} & {\bf 0.3}\\
         {\bf 0.6} & {\bf 0.1}
         \end{array}
         \right)$,
 $\bf{w}_1= \left(\begin{array}{cc}
         {\bf 0.9} & {\bf 0.8}\\
         {\bf 0} & {\bf 1}
         \end{array}
         \right)$, ${\bf b}_0 = \bf 0.1$, ${\bf b}_1 =\bf 0.15$.
\end{center}

 The expected output $y = \bf{0.8}$, the admitted error $\varepsilon = \bf{10^{-1}}$, the learning rate $\eta = \bf 0.1$. Note that $\hat{y}={\bf \lambda}_1\vee {\bf \lambda}_2$, according to our choice for aggregating the output values.

In the following, the training process of our simple neural network is performed as a sequence of inferences in our system.

\medskip

\begin{tabular}{llr}
(1) & $ [init({\bf h})]\cf{({\bf w}_0, {\bf w}_1),
 ({\bf b}_0,{\bf b}_1)}\to [train({\bf h})]\cf{({\bf w}_0, {\bf w}_1),
 ({\bf b}_0,{\bf b}_1)}$ & (N0) \\ 
(2) & $[train({\bf h})]\cf{({\bf w}_0, {\bf w}_1),
 ({\bf b}_0,{\bf b}_1)}\to [train(next_{{\bf w}_0,{\bf b}_0}({\bf h}))]\cf{{\bf w}_1,{\bf b}_1}$ & (N1) 
 \end{tabular} 
\medskip
 
\noindent If $ {\bf a}^1 =(a_1^1, a_2^1)=next_{{\bf w}_0,{\bf b}_0}({\bf h})$, then ${\bf a}^1= (\bf{0.27}, \bf{0.25})$. 
\medskip

 \begin{tabular}{llr}
 (3) & $[train({\bf a})]\cf{{\bf w}_1,{\bf b}_1}\to [train(next_{{\bf w}_1,{\bf b}_1}({\bf a}))]\cf{}$ & (N1)
  \end{tabular}
  
\medskip
  
\noindent We note that  $ {\bf \lambda} =({\bf \lambda}_1, {\bf \lambda}_2)=next_{{\bf w}_1,{\bf b}_1}({\bf a})= (\bf{0.393}, \bf{0.626})$. 
 
 \medskip
 
 \begin{tabular}{llr}
 (4) & $ [init({\bf h})]\cf{({\bf w}_0, {\bf w}_1),
 ({\bf b}_0,{\bf b}_1)}\to [train({\bf \lambda})]\cf{}$ & (1,2,3)
  \end{tabular}  
\medskip
  
\noindent We note that $\hat{y} = \bf{0.626}$, so  $end({\bf y}, 
 {\bf \lambda},{\bf \varepsilon})= d_L({\bf y},  { \hat{y}})\ral \bf{\varepsilon}$ is equivalent with $\bf{0.174}\ral \bf{0.1}$, which means that    $@^{ln}_{1_L}end({\bf y},  {\bf \lambda}, {\bf \varepsilon})$ is {\em false}. Consequently, we apply (N2):
\medskip

 \begin{tabular}{llr}
 (5) & $ [init({\bf h})]\cf{({\bf w}_0, {\bf w}_1),
 ({\bf b}_0,{\bf b}_1)}\to ([train({\bf \lambda})]\cf{}\wedge 
 \neg@^{ln}_{1_L}end({\bf y},  {\bf \lambda}, {\bf \varepsilon})\to $ & \\
 &\hspace*{1cm}$\to [init({\bf h})]updated_{{\bf \lambda}}\cf{({\bf w}_0, {\bf w}_1),
 ({\bf b}_0,{\bf b}_1)})$ & (N2)
  \end{tabular}

\medskip

\noindent We have to  update the weigths and biases. In order to perform this computation, we use the Gradient Descent formula, so 
$updated_{\bf\lambda}\cf{({\bf w}_0, {\bf w}_1),
 ({\bf b}_0,{\bf b}_1)}=\cf{({\bf uw}_0, {\bf uw}_1),
 ({\bf ub}_0,{\bf ub}_1)}$ such that:

\begin{center}
(GD) $uw_{ij}^0 := ReLU_1(
w_{ij}^0 - \eta \frac{\partial d_L(\bigvee_1^2 \lambda_i, \hat{y})}{\partial w_{ij}^0})$  and $uw_{ij}^1 := ReLU_1( 
w_{ij}^1 - \eta \frac{\partial d_L(\bigvee_1^2 \lambda_i, \hat{y})}{\partial w_{ij}^1})$ 
\end{center}

\noindent and similar updates are performed on biases.  The new (updated)  weights and biases are: 
\begin{center}
 $\bf{uw_0}= \left(\begin{array}{cc}
         {\bf 0.416} & {\bf 0.324}\\
         {\bf 0.62} & {\bf 0.13}
         \end{array}
         \right)$,
 $\bf{uw_1}= \left(\begin{array}{cc}
         {\bf 0.9} & {\bf 0}\\
         {\bf 0.827} & {\bf 1}
         \end{array}
         \right)$,                 
${\bf ub}_0 = \bf 0.28$, ${\bf ub}_1 =\bf 0.25$.
\end{center}

\noindent Consequently, the second trainig step is performed by:
\medskip

\begin{tabular}{llr}
(6) & $ [init({\bf h})]\cf{({\bf uw}_0, {\bf uw}_1),
 ({\bf ub}_0,{\bf ub}_1)}\to [train({\bf h})]\cf{({\bf uw}_0, {\bf uw}_1),
 ({\bf ub}_0,{\bf ub}_1)}$ & (N0) 
 \end{tabular}

\medskip
\noindent and the logical training  continues until  (N3)  can be applied. The use of $ReLU_1$ for each backward step implies that the  updated weights are definable in our logic.

Regarding our decisions and the formalization of the training process we note the following.

\begin{itemize}

\item In axioms (N2) and (N3) we analyze the error in order to take a decision. By (n2),  the output is calculate as ${\bigvee}_1^n{\bf \lambda}_i$  where $\lambda_i$ defined the $i^{th}$ output after the last hidden layer (consequently, the same formula is used in (GD)). We can consider other aggregators but note that, in our approach, they should be defined as formulas in \L ukasiewicz logic with rational coefficients.

\item  The logical system  system ${\mathcal H}_{\bm \Sigma}(@)+\Lambda_{MLP}$ can be adapted for verifying network properties.  In the following we show that we can  track the number of epochs  of the training process and  abandon the training if the desired outcome is not reached within the given limit. Let  $E\in \N $ be our limit. If $1_E=1/E$ then $1_E\oplus\ldots\oplus 1_E=1$ where the \L ukasiewicz sum has $E$ terms, so we can control   the number of epochs as an $rmv$-formula. This formula will be the first argument of the configuration operator $\cf{\su }:rmv^n\to ln$. When a training epoch is initiated, this number is checked and the training stops if the maximum is reached (see axiom (N0$_E$) below).  The new axioms are:

\begin{itemize}

\item[(N0$_{\neg E}$)]  $[init ({\bf h}_1^n)]\cf{{\bf r}, {\bf w}_0^k, {\bf b}_0^k}\wedge \neg @^{ln}_{1_L}{\bf r}\to [train ({\bf h}_1^n)]\cf{{\bf r},{\bf w}_0^k, {\bf b}_0^k}$\\

\item[(N0$_{E}$)]  $[init ({\bf h}_1^n)]\cf{{\bf r}, {\bf w}_0^k, {\bf b}_0^k}\wedge @^{ln}_{1_L}{\bf r}\to [stop()]
\cf{{\bf 1}_L,{\bf w}_0^k, {\bf b}_0^k}$\\

\item[(N1)] $[train ({\bf h}_1^n)]\cf{{\bf r},{\bf w}_i^k, {\bf b}_i^k}\to [train (next_{{\bf w}_i, {\bf b}_i}({\bf h}_1^n))]\cf{{\bf r},{\bf w}_{i+1}^k, {\bf b}_{i+1}^k }$\\

\item[(N2)] $[init ({\bf h}_1^n)]\cf{{\bf r},{\bf w}_0^k, {\bf b}_0^k} \to ([train(\lambda_1^n)]\cf{}\wedge \neg @^{ln}_{1_L}end({\bf y},  \lambda_1^n, {\bf \varepsilon})\to$\\
\hspace*{6cm} $ [init ({\bf h}_1^n)]
updated_{\lambda_1^n}\,\,  \cf{{\bf r}\oplus 1_E, {\bf w}_0^k, {\bf b}_0^k})$\\

\item[(N3)] $[init ({\bf h}_1^n)]\cf{{\bf r}, {\bf w}_0^k, {\bf b}_0^k} \to ([train(\lambda_1^n)]\cf{{\bf r}}\wedge  @^{ln}_{1_L}end({\bf y},  \lambda_1^n, {\bf \varepsilon})\to [stop (\lambda_1^n)]
\cf{{\bf r},{\bf w}_0^k, {\bf b}_0^k})$
\end{itemize}


\end{itemize}

The system ${\mathcal H}_{\bm \Sigma}(@)+\Lambda_{MLP}$ is implemented and tested in the Lean 4 proof assistant and the implementation is briefly explained in Section \ref{lean}. 
In Lean 4 we perform:

\begin{itemize}
\item numerical computation: we have implemented the training algorithm, taking advantage of Lean 4 capabilites as programming language;
\item logical execution: the training process is a sequence of logical inferences; in Section \ref{loginf}  we define a concrete network and its training  with a limited number of epochs as logical deduction; moreover, once the network has been trained, we can extract the logical formula associated with it (see Section \ref{loginf});
 
\item model generation:  each state of the model corresponds to the current state of the network; the main steps of the algorithm are presented in Section \ref{model}, each step being determined by a logical inference that also represents  a step in the training process.  
\end{itemize}

We provided this example to illustrate the mechanism of logical inference, and a real-world model is presented in Section \ref{section:experiment}.

\subsection{\change{Backpropagation}}

In our framework, backpropagation is formulated entirely within \L ukasiewicz logic, as every stage of the training process is computed in the unit interval $[0,1]$ and employs only the MV-algebraic operations.  

For each layer $t \in 1 \dots k$, the forward pass is defined by 
$\mathbf{a}_t := ReLU_1(\mathbf{z}_{t})$, where  
$\mathbf{z}_{t} = \mathbf{w}_{t} \mathbf{a}_{t-1} + \mathbf{b}_{t}$. The derivative of the activation is represented as a diagonal matrix $D_{t} := \mathrm{diag}\!\big(\mathbf{1}_{(0,1)}(z_{t_1}), \ldots, \mathbf{1}_{(0,1)}(z_{t_{n_t}})\big)$, 
where $n_t$ is the number of neurons of the $t$$^{th}$ layer and $\mathbf{1}_{(0,1)}(z)$ is the characteristic function: $\mathbf{1}_{(0,1)}(z) = 1$ if $0 < z < 1$ and $0$ otherwise. We note that at the output layer the initial gradient is  $\mathbf{g} := \mathrm{sign}(\mathbf{a}_{k} - \mathbf{y}) \in \{-1,0,1\}^{n_k}$. 

The loss is measured using the \L ukasiewicz distance $d_L$, and backpropagation proceeds by the chain rule. For any hidden layer $t$, the error is propagated as $\nabla_{\mathbf{z}_{t}} d_L = \Pi_{t} \mathbf{g} \in \mathbb{R}^{n_t}$, where
\begin{center}
$\Pi_{t} := D_{t} \mathbf{w}^{\top}_{t+1} D_{t+1} \mathbf{w}^{\top}_{t+2} \cdots D_{k}$.
\end{center} 
The parameter gradients follow as $G_{\mathbf{w}_t} =\nabla_{\mathbf{w}_{t}} d_L = \nabla_{\mathbf{z}_{t}} d_L \, \mathbf{a}^{\top}_{t-1} \in \mathbb{R}^{d_t \times d_{t-1}}$, and $G_{\mathbf{b_t}} = \nabla_{\mathbf{b}_{t}} d_L = \nabla_{\mathbf{z}_{t}} d_L \in \mathbb{R}^{d_t}$.

Since raw gradients may lie outside $[0,1]$, they are normalized by the $\ell_\infty$-norm:
\begin{equation*}
\hat{\mathbf{g}} = \frac{|\mathbf{g}|}{\|\mathbf{G}\|_\infty + \varepsilon}, 
\qquad \hat{\mathbf{g}} \in [0,1] \text{ and } \varepsilon > 0,
\end{equation*}
ensuring compatibility with \L{}ukasiewicz operations. The learning rate $\eta \in [0,1]$ is then combined with these magnitudes via the \L{}ukasiewicz product, $\Delta = \eta \otimes \hat{\mathbf{g}}$. Finally, the parameter update is expressed exclusively in \L ukasiewicz terms. For each weight, $\mathbf{uw}_{k} = \big(\mathbf{w}_{k} \ominus \Delta^- \big) \oplus \Delta^+$, where
\[
\Delta^+ =
\begin{cases}
\eta \otimes \hat{\mathbf{g}}, & g < 0, \\
0, & \text{otherwise},
\end{cases}
\qquad
\Delta^- =
\begin{cases}
\eta \otimes \hat{\mathbf{g}}, & g > 0, \\
0, & \text{otherwise}.
\end{cases}
\]

Thus, every update rule is internalized into \L ukasiewicz operations $(\oplus,\ominus,\otimes)$, ensuring that the backpropagation process remains entirely inside the algebraic-logical framework.

\section{Implementation in Lean}\label{lean}
To define the Lean \cite{lean4} implementation of this system, we rely on several key modules: the implementation of MV-algebras and DMV-algebras as extensions of abelian monoids; the implementation of linear algebra operations over dependent types for vectors and matrices; the implementation of the Multi-Layer Perceptron; finally, the implementation of a Many-Sorted Hybrid Logic-both its language and its proof system-which integrates all the components mentioned above.

With these modules in place, we implement an algorithm that generates a model capturing the execution of the neural network within the logical system.

The full implementation is available at 

\url{https://github.com/bogdanmacovei/lukasiewicz-neural-network-formalization}.

\subsection{MV-algebras and rational coefficients}

We begin by presenting the definition of the algebraic structure corresponding to an MV-algebra. The complete implementation, including related structures, is provided in the source code. An MV-algebra extends the underlying abelian monoid by incorporating three additional axioms, defined as follows:

\begin{lstlisting}[language=Lean, numbers=none,basicstyle=\footnotesize, frame=none]
structure MV (α : Type) extends AbelianMonoid α := 
    (neg : α → α) (mul : Float → α → α)
    (MV₂ : ∀ x : α, neg (neg x) = x)
    (MV₃ : ∀ a : α, add (neg zero) a = neg zero)
    (MV₄ : ∀ x y : α, add (neg (add (neg x) y)) y = add (neg (add (neg y) x)) x)
\end{lstlisting}

In a similar manner, the definition of a DMV-algebra builds upon the existing MV-algebra structure by incorporating an additional set of axioms. For clarity, we illustrate the definition with a representative example, as the remaining axioms follow an analogous pattern and do not introduce notable implementation challenges.

\begin{lstlisting}[language=Lean, numbers=none, basicstyle=\footnotesize, frame=none]
structure DMV (α : Type) extends MV α := 
    (r_values : interval01)
    (DMV₁ : ∀ r : Float, ∀ x y : α, List.elem r r_values.list → mul r x = mul r y)
\end{lstlisting}

For the linear algebra component, we define two dependent data types, \texttt{Vector} and \texttt{Matrix}. These structures play a central role, as they are employed both in the algorithmic execution of the network and in the specification of the proof system. All operations related to training and weight updates are defined in terms of these data types.

\begin{lstlisting}[language=Lean, numbers=none, basicstyle=\footnotesize, frame=none]
def Vector (α : Type) (n : Nat) := Fin n → α   
def Matrix (α : Type) (m n : Nat) := Fin m → Vector α n 
\end{lstlisting}

We introduce the following notations: \texttt{v[i]} to access the \texttt{i}$^{\text{th}}$ element of a vector \texttt{v}, and \texttt{m[i, j]} to access the \texttt{j}$^{\text{th}}$ element of the \texttt{i}$^{\text{th}}$ row of a matrix \texttt{m}.

\subsection{${\mathcal H}_{ \Sigma}(@)+\Lambda_{MLP}$}\label{leanlog}
We are now prepared to define the language of our many-sorted hybrid logic. we introduce inductive types for sorts, actions, and formulas, which provide the necessary foundations for specifying the axioms of the system. Moreover, these types enable the logical computation performed by the network.
\begin{multicols}{2}
\begin{lstlisting}[language=Lean, numbers=none, basicstyle=\footnotesize, frame=none]
inductive ActionNN (σ : Nat)
| atom : Fin σ → ActionNN σ
| init : ActionNN σ
| train : ActionNN σ
| update : ActionNN σ
| Stop : ActionNN σ
| seq : ActionNN σ → ActionNN σ → ActionNN σ
\end{lstlisting}

\bigskip
\bigskip

\begin{lstlisting}[language=Lean, numbers=none, basicstyle=\footnotesize, frame=none]
inductive FormNN (σ : Nat) : Type where
| atom : Fin σ → FormNN σ
| neg : FormNN σ → FormNN σ
| oplus : FormNN σ → FormNN σ → FormNN σ
| sigma : Set (FormNN σ) → FormNN σ
| nom : Nominal σ → FormNN σ → FormNN σ
| diam : Nominal σ → FormNN σ → FormNN σ
| bot : FormNN σ
\end{lstlisting}
\end{multicols}

We show the implementation of (N0)-(N3) axioms, formally introduced in the prevous section as part of our logical framework.

\begin{lstlisting}[language=Lean, numbers=none, basicstyle=\footnotesize, frame=none]
N0 {Γ : Ctx σ} {φ ψ : FormNN σ} : ProofNN Γ $ [[ActionNN.init]] φ ⊃ [[ActionNN.train]] ψ
N1 {Γ} {n m : Nat} {W : Matrix Float n m} {b : Float} {ϕ : List (FormNN σ)} :
    ProofNN Γ $ [[ActionNN.train]] (FormNN.list ϕ) ⊃ FormNN.list (layer_activation_form b W ϕ)
N2 {Γ : Ctx σ} {n k : Nat}
     {W : Vector (Matrix Float n n) k} {b : Vector Float k}
     {input : Vector Float n} {L : List (FormNN σ)}
     {target : FormNN σ} {ε : FormNN σ} :
    let ψ := FormNN.list (encode_pair W b)
    let trainPart := [[ActionNN.train]] (FormNN.list L)
    let condition := ¬L (FormNN.hybrid (#n 1) (sort.atom 0) (target L ε))
    ProofNN Γ $ ([[ActionNN.init]] ψ ⊃ (trainPart & condition)) ⊃ [[ActionNN.update]] ψ
N3 {Γ : Ctx σ} {n k : Nat}
     {W : Vector (Matrix Float n n) k} {b : Vector Float k}
     {input : Vector Float n} {L : List (FormNN σ)}
     {target : FormNN σ} {ε : FormNN σ} :
    let ψ := FormNN.list (encode_pair W b)
    let trainPart := [[ActionNN.train]] (FormNN.list L)
    let condition := FormNN.hybrid (#n 1) (sort.atom 0) (target L ε)
    ProofNN Γ $ ([[ActionNN.init]] ψ ⊃ (trainPart & condition)) ⊃ [[ActionNN.Stop]] ψ
\end{lstlisting}

With all the necessary language constructs in place, we can now formally describe the training algorithm of the network. We proceed by defining, in an iterative manner, the activation of a single neuron, the activation of a layer, and ultimately the training of the entire network. Due to the repetitive nature of the definitions, we focus here on illustrating the activation of one layer. The training of the full network is then obtained by composing successive layer activations, from the input layer to the output layer. This sequence of activation and backpropagation steps yields the automatically generated model of the network: a symbolic execution trace that starts from an initial configuration and unfolds as a sequence of intermediate states, that has as the final state the trained version of the network. 

\begin{lstlisting}[language=Lean, numbers=none, basicstyle=\footnotesize, frame=none]
def forwardLayer {n : Nat} (w : Matrix n n) (b : Vector n) (input : Vector n) : Vector n :=
    fun i => let sum := (List.finRange n).foldl (fun acc j => acc + w i j * input j) 0
    relu (sum + b i)
\end{lstlisting}

The backpropagation is defined by a set of recursive equations that update the weights and biases of the network based on the error propagated from the output layer backwards. These updates rely on the gradient of the loss function with respect to each parameter, and are expressed over the matrix and vector structures previously introduced. 

\begin{lstlisting}[language=Lean, numbers=none, basicstyle=\footnotesize, frame=none]
def updateStep {n k : Nat} (s : NNState n k) : NNSTATE n k := 
    let idx := s.id + 1, newBiases := fun l => fun i => 
    let sum := List.finRange n |>.foldl (fun acc j => acc + s.weights l i j * s.input j) 0.0
    let act := relu1 sum; s.biases l i - learning_rate * act 
    { id := idx, weights := s.weights, biases := newBiases, input := s.input }
\end{lstlisting} 

\subsection{Example of logical inference}\label{loginf}

Using the axioms, we can derive more advanced properties about the training process; in particular, we can formally prove that if an update step is being executed, the network has not yet reached the final epoch. This corresponds to the fac that the error remains above the given threshold and that the gradient descent is still contributing meaningfully to weight adjustments. The following theorem captures this idea formally and expresses it in our logic:

\begin{lstlisting}[language=Lean, numbers=none, basicstyle=\footnotesize, frame=none]
theorem inductive_step_termination
    {n m k : Nat} {y η ε E : Float} {Γ : Ctx σ}
    {W : Vector (Matrix Float n m) k} {b : Vector Float k}
    {lφ : List $ FormNN σ} {ln : sort σ}
    {mem ν : FormNN σ}
    [Inhabited $ FormNN σ] [Inhabited $ Nominal σ] [OfNat (Fin σ) 0] :
    Γ ⊢ ([[ActionNN.train]] ⟪mem, ν⟫ ⊃ FormNN.list lφ) →
    Γ ⊢ [[ActionNN.update]] ⟪mem ⊕ nomToForm (#γ (1/E)), ν⟫ →
    Γ ⊢ ~@@(#n 1), ln : ((dL (nomToForm (#γ y)) (foldr (fun φ ψ => φ ⋁ ψ) zL lφ)) →L nomToForm (#γ ε)) →
    Γ ⊢ (@@(#n 1), ln : dL mem (nomToForm (#γ ((E - 1)/E)))) &
        [[ActionNN.train]] ⟪mem ⊕ nomToForm (#γ (1/E)), ν⟫
\end{lstlisting} 

Upon completion of training, the network yields two complementary outputs, a concrete numerical result and a logical formula. To illustrate this dual outcome, we consider a simple experiment of training a network with three layers: an input layer with two nodes, a hidden layer with three nodes, and the output layer with one node. We write this information in Lean:

\begin{lstlisting}[language=Lean, numbers=none, basicstyle=\footnotesize, frame=none]
def inputs : List $ form σ := [nomToForm $ #γ 0.2, nomToForm $ #γ 0.3]
def w1 : List $ List Float := [[0.1, 0.2], [0.13, 0.03], [0.11, 0.1]]
def w2 : List $ List Float := [[0.3, 0.67, 0.15]]
def b : List $ Float := [0.08, 0.18]
\end{lstlisting}

The training process yields two distinct results, a numerical output (0.222), reflecting the evaluation of the network on the give input, and a formula that encodes the corresponding logical structure of the trained network:

\begin{lstlisting}[language=Lean, numbers=none, basicstyle=\footnotesize, frame=none]
((γ0.18) ⊕ ((◇0.30 ((γ0.08) ⊕ ((◇0.10 (γ0.20) ⊕ ((◇0.20 (γ0.20) ⊕ ((◇0.00 (γ0.20) ⊕ (γ0.00))))) ⊕ (γ0.00)))
\end{lstlisting}

\subsection{Automatically Generated Network Execution Model}\label{model}

Having defined the algebraic structures, the neural network architecture, and the logical framework, we now introduce the symbolic construction of the model corresponding to a training execution. This model captures the full sequence of states traversed by the network during its training process, using the formalism introduced in our logic.

We begin from an initial state \texttt{init}, which encodes the initial weights and biases. From this state, we iteratively apply actions defined by the logic, \texttt{train}, \texttt{updated} and \texttt{stop}, to generate transitions representing the evolution of the system.

Each training step corresponds to a forward action, and each update step modifies the weights and biases according to the output of the network at that stage. The process is repeated until the computed loss, obtained via a symbolic forward pass and a loss function, falls below a threshold.

Formally, the training model is generated using the function:

\begin{lstlisting}[language=Lean, numbers=none, basicstyle=\footnotesize, frame=none, mathescape=false]
def symbolicTrainLoop {n k : Nat} (init : NNState n k) (target : Vector' n) (threshold : Float) : List (Transition n k)
\end{lstlisting}

This function repeatedly executes training and update actions, storing each transition in a list. The algorithm proceeds as follows:

\begin{enumerate}
\item Start from the initial state $s_0$, with initial weights and biases
\item apply \texttt{Action.train} to compute a new state via forward propagation
\item Evaluate the output of the network.
\item Compute the loss with respect to the given target vector.
\item \textbf{If} the loss is below the given threshold: (5.1) apply \texttt{Action.stop} to finalize the training and (5.2) terminate the algorithm and return the list of all transitions and the final state, with the computed weights and biases.
\item \textbf{Else}: (6.1) apply \texttt{Action.update} to adjust the biases and (6.2) repeat from step 2 for the next epoch, up to the maximum allowed number of epochs.
\end{enumerate}

\subsection{\change{Experiment}}
\label{section:experiment}

In order to complement the theoretical framework and the formalization of this system, we implement and train a multiplayer perceptron on a standard dataset, consisting of the two-moons binary classification benchmark, with 6000 training examples and 2000 test examples, balanced across the two classes. Each input vector is two-dimensional and scaled to the unit interval $[0,1]$. 

We use a fully connected architecture with two hidden layers of 32 neurons each, followed by a single output unit. All activations are defined using $ReLU_1$ function and a learning rate $\eta = 1$, with all operations expressed entirely in \L ukasiewicz arithmetic. We use in the training process mini-batches of size 128 for 250 epochs. For comparison, we also train a classical Python baseline with the same architecture, but using the standard $ReLU$ in the hidden layers, and a \textit{sigmoid} output, binary cross-entropy and stochastic gradient descent as the optimization algorithm. 

The results are summarized in Table \ref{tab:results}. The Lean \L ukasiewicz MLP achieves stable learning, reaching around 90\% training accuracy and 89\% test accuracy, showing that meaningful classification can be performed even under the strict arithmetic constraints. The classical baseline achieves 96\% accuracy on both train and test sets. These results suggest that further refinements such as smoother losses or fuzzy cross-entropy could narrow the gap between these two models.

\begin{table}[h]
    \centering
    \begin{tabular}{|c|c|c|}
        \hline 
        \textbf{Model} & \textbf{Train Accuracy} & \textbf{Test Accuracy} \\ 
        \hline 
        Lean \L ukasiewicz MLP & 0.9 & 0.89 \\  
        \hline 
        Python Classic MLP & 0.96 & 0.96 \\ 
        \hline 
    \end{tabular}
    \caption{Comparative results}
    \label{tab:results}
\end{table}

\section{Related work and conclusions}\label{related}

  The present paper is a contribution towards defining and analyzing a neural network within a logical framework, an active direction of research, that can presumably 
 "assist in preventing unfair, unwarranted, or otherwise
undesirable outcomes from the application of modern AI 
methods" \cite{lnn2}.

The role of fuzzy logic in Artificial Intelligence is well-known. We follow \cite{hajek}, where fuzzy logics are defined as $t$-norm based logics on the real (rational) interval $[0,1]$.  The evolution of \L ukasiewicz logic and its corresponding algebraic theory was presented in the preliminary section, but we have to mention that the idea of representing neural networks as formulas of  an extension of \L ukasiewicz logic goes back to \cite{Trillas}. The recent theory of Logic Neural Networks \cite{lnn1,lnn2} is a  major step forward towards a systematic treatment of neural networks using  various t-norm based  logics. 
 
Hybrid modal logic is a well-established research topic in modal logic, rooted in the work of Arthur N. Prior on tense logic. We refer to \cite{hybtemp} for a comprehensive study.    Our setting is based on the general many-sorted hybrid modal logic from \cite{tableaux}, where the system was used for representing a (toy) programming language and its operational semantics. A complex system of many-sorted hybrid modal logic was also proposed in \cite{tutu}, with applications to biological processes.

In the past years, formal verification emerged as a useful technique  for certifying the behavior of a neural network. One can see  \cite{verif2} for an overview on this topic. We especially mention  the framework NeSAL \cite{verif1} that proposes a Hoare-like logic for neural networks. In \cite{tableaux}, the system ${\mathcal H}_{\Sigma}(@,\forall)$ (which is an extension of ${\mathcal H}_{\Sigma}(@)$ with quantification for variables interpreted as singletons) was powerful enough to model an programming language, its semantics and an adequate version of Hoare logic for verification. This result gives us hope that, in the future, we can will be able to define  a  framework similar with  NeSAL  within our logic.

Our results on logical representation of neural network seems to be promising, at least in our particular setting: we used a general system of hybrid modal logic in which a multilayer perceptron with $ReLU_1$ activation function is defined as a particular theory. As explained in Section \ref{mlp},  our choice for Lean 4  is motivated by its dual nature: "an extensible theorem prover and an efficient programming language"  \cite{lean4}. Since the formal proofs are complex, we have tried to lower the complexity of our logic. Consequently, we did not add the full power of equational reasoning  in representing \L ukasiewicz logic with rational constants.  However, this can be done using more elaborate  systems of equational reasoning for hybrid modal logic \cite{eq1, eq2, eq3}, and  the Lean automatic inference can be better used defining appropriate tactics.

In conclusion, we propose many-sorted hybrid modal logic as a general and expressive system that, through particular choices of sorts and operations,  can be used to define a logical theory for the training process of a neural network. Lean 4 provides support for  specification, verification and  algorithmic training.

\end{document}